\documentclass[nofootinbib,aps,prb,twocolumn,superscriptaddress,amssymb,floatfix,longbibliography]{revtex4-2}

\makeatletter
\def\pdfauthorlist{}
\let\revtex@author\author
\renewcommand{\author}[1]{%
  \revtex@author{#1}%
  \ifx\pdfauthorlist\@empty
    \gdef\pdfauthorlist{#1}%
  \else
    \g@addto@macro\pdfauthorlist{, #1}%
  \fi
}
\makeatother

\newcommand{\GOEaffiliation}{\affiliation{Institut für Theoretische Physik, Georg-August-Universität Göttingen, Friedrich-Hund-Platz 1, 37077 Göttingen, Germany}}
\newcommand{\PSUaffiliation}{\affiliation{Department of Physics, The Pennsylvania State University, University Park, 16802 Pennsylvania, USA}}

\usepackage{xcolor}
\usepackage{graphicx,upgreek}
\usepackage{hyperref}
\usepackage[all]{hypcap}
\usepackage{mathtools}
\usepackage{physics}
\usepackage{bbm}
\usepackage{orcidlink}

\usepackage{lmodern}
\usepackage{microtype}

\usepackage[utf8]{inputenc}
\usepackage[T1]{fontenc}

\usepackage[english]{babel}
\makeatletter
\adddialect\l@en\l@english
\makeatother

\newif\ifqophat
\qophattrue 

\newcommand{\qop}[1]{%
  \ifqophat
    \hat{#1}%
  \else
    #1%
  \fi
}

\renewcommand{\paragraph}[1]{{\bfseries #1}---\!\!}

\makeatletter
\hypersetup{
    colorlinks=true,
    linkcolor=blue,
    citecolor=blue,
    filecolor=blue,
    urlcolor=blue,
}
\makeatother

\begin{document}

\title{Random-matrix and transport frequencies in eigenstate spectral functions}

\author{Kadir Çeven\orcidlink{0000-0002-1770-1255}}
\GOEaffiliation

\author{Rohit Patil\orcidlink{0009-0005-8250-2382}}
\PSUaffiliation

\author{Marcos Rigol\orcidlink{0000-0002-5806-5873}}
\PSUaffiliation

\author{Fabian Heidrich-Meisner\orcidlink{0000-0002-3463-1121}}
\GOEaffiliation

\date{\today}

\begin{abstract}
The fact that generic isolated many-body quantum systems thermalize is understood using the eigenstate thermalization hypothesis (ETH). In recent years, there has been much interest in the behavior of the ETH spectral functions, which characterize the smooth dependence of the variance of the off-diagonal matrix elements of observables on the associated energy and frequency, and whose low-frequency part contains information about the long-time dynamics. In finite systems described by the ETH, the spectral functions are expected to exhibit plateaus below a characteristic frequency $\omega^{}_{\mathrm{ETH}}$. In this regime, the statistics of the matrix elements of observables are expected to be described by random matrix theory. Related frequencies that have been studied in the literature are $\omega^{}_{\mathrm{SFF}}$, which controls the onset of random-matrix behavior in the spectral form factor, and the transport frequencies $\omega^{}_{\mathrm{tr}}$, which are derived from transport coefficients. However, a direct quantitative comparison of these frequencies is lacking. Using exact diagonalization, we conduct such a comparison for the spectral functions of current operators in clean and disordered quantum spin ladders with diffusive energy and spin transport. We find clear evidence for the expected low-frequency plateaus in the ETH spectral functions. For the accessible system sizes, $\omega^{}_{\mathrm{SFF}}$ is consistent with the extent of the plateaus, while the transport frequencies are systematically larger and lie in the nonuniversal regime of the spectral functions. Our findings highlight the need to better understand the origin of these quantitative differences.
\end{abstract}

\maketitle

\makeatletter
\hypersetup{pdfauthor=\pdfauthorlist, pdftitle=\@title}
\makeatother

\section{Introduction}\label{sec:intro}

A central question in the nonequilibrium physics of closed many-body quantum systems is how thermalization emerges from unitary quantum dynamics~\cite{DAlessio2016, Gogolin2016, Mori2018, Patil2026A}. In a nutshell, generic many-body Hamiltonians behave in many ways like random matrices drawn from appropriate ensembles~\cite{DAlessio2016}. This picture is formalized by the eigenstate thermalization hypothesis (ETH)~\cite{Deutsch1991, Srednicki1994, Srednicki1999, Rigol2008, DAlessio2016, Patil2026A}, which generalizes the random matrix theory (RMT) prediction for the first two moments of the distribution of the matrix elements of Hermitian operators to few-body observables in local many-body quantum Hamiltonians (see also Refs.~\cite{Foini2019, Foini2019A, Pappalardi2022, Pappalardi2025, Vallini2026} for the extension to higher moments). The validity of the ETH, which has been verified in numerous examples (see, e.g.,~Refs.~\cite{Rigol2008, Rigol2009a,*Rigol2009b, Rigol2010, Steinigeweg2014B, Beugeling2015, Mondaini2017, Jansen2019, LeBlond2019, LeBlond2020, Schonle2021, Vidmar2021, Patil2025, Swietek2025, Patil2026, Swietek2026}), implies thermalization under broad conditions on the initial state. 

\begin{figure}
\includegraphics[width=\linewidth]{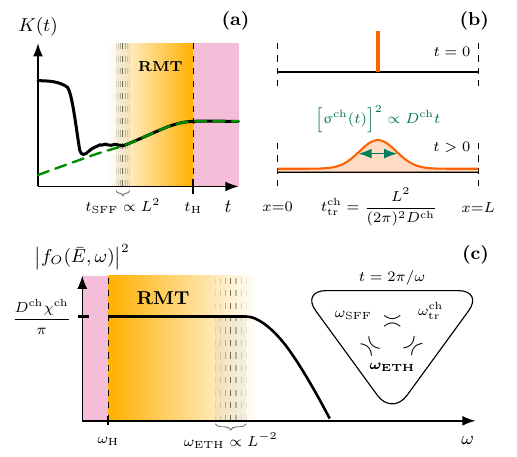}
\vspace{-0.5cm}
\caption{Illustration of the quantities investigated in this work: (a) The spectral form factor $K(t)$, (b) transport coefficients such as diffusion constants, which are related to the spreading of a density perturbation, (c) the spectral function $\abs{f^{}_O(\bar{E}, \omega)}^2$. (a) In the SFF, RMT behavior sets in at the time $t^{}_\mathrm{SFF}$ where $K(t)$ starts to increase linearly in time. This corresponds to a frequency $\omega^{}_{\mathrm{SFF}}=2\pi/t^{}_\mathrm{SFF}$. (b) The transport time $t_\mathrm{tr}^{\mathrm{ch}}=2\pi/\omega^{\mathrm{ch}}_{\mathrm{tr}}$ describes the scaling with $L$ of the time at which a perturbation in a density spreads across the system. $\upsigma^{\mathrm{ch}}(t)$ is the width of such a density perturbation. (c) In the spectral function $|f^{}_O(\bar{E}, \omega)|^2$ defined through the ETH ansatz in Eq.~\eqref{eq:ETH_ansatz}, a plateau accompanied by RMT behavior is expected for $\omega^{}_{\mathrm{H}} < \omega \lesssim \omega^{}_{\mathrm{ETH}}$, where $\omega^{}_{\mathrm{H}} = 2\pi/t^{}_{\mathrm{H}}$. We study the ETH for current operators. Therefore, $\abs{f^{}_O(\bar{E}, \omega \rightarrow 0)}^2 = D^{\mathrm{ch}} \, \chi^{\mathrm{ch}}/\pi$, where $\chi^{\mathrm{ch}}$ is the leading coefficient in the high-temperature expansion of the associated static susceptibility.}
\label{fig:schematic}
\end{figure}

In a generic isolated many-body system, the time window over which spectral RMT behavior occurs depends on the system size. This regime is characterized by $t^{}_{\mathrm{SFF}}(L) \lesssim t \lesssim t^{}_{\mathrm{H}}(L)$, within which the spectral form factor (SFF) $K(t)$ agrees with RMT predictions [see the sketch in Fig.~\ref{fig:schematic}(a)]. Here, $t^{}_{\mathrm{SFF}}$ defines the onset of RMT behavior in the SFF~\cite{Chan2018, Bertini2018}, while $t^{}_{\mathrm{H}}$ is the Heisenberg time, which is determined by the inverse level spacing at the corresponding energy. $t^{}_{\mathrm{SFF}}$ grows polynomially with the linear system size $L$ in diffusive, sub- and superdiffusive systems, inheriting its $L$-dependence from the slowest transport channel, while $t^{}_{\mathrm{H}}$ grows exponentially with $L$~\cite{Kos2018, Gharibyan2018, Sierant2020B, Roy2020, Suntajs2020, Colmenarez2022, Roy2022, Kumar2024, Kumar2026}. If $t^{}_{\mathrm{SFF}}$ is comparable to $t^{}_{\mathrm{H}}$, RMT behavior can fail to occur, signaling a possible breakdown of thermalization~\cite{Suntajs2020, Suntajs2022, Kliczkowski2023, Sierant2025, Swietek2025, Swietek2026}.

In the context of transport, the longest characteristic time scale associated with a given channel is the transport time $t^{\mathrm{ch}}_{\mathrm{tr}}$, set by the longest wavelength in the system~\cite{Edwards1972, Thouless1974, Thouless1977}. For a diffusive system with periodic boundary conditions, $t^\mathrm{ch}_{\mathrm{tr}}$ characterizes the relaxation of the slowest hydrodynamic mode, $t^{\mathrm{ch}}_{\mathrm{tr}} \simeq [L/(2\pi)]^2/D^\mathrm{ch}$, where $D^\mathrm{ch}$ is the corresponding diffusion constant~\cite{Thouless1974, Thouless1977}. Up to a nonuniversal prefactor, $t^{\mathrm{ch}}_{\mathrm{tr}}$ can be understood as the time needed for a local perturbation in the corresponding density to spread across a finite system [see Fig.~\ref{fig:schematic}(b)].

Another way to identify the regime over which RMT behavior occurs is through the ETH spectral function. The ETH ansatz~\cite{Srednicki1999} posits that the matrix elements $O_{\alpha\beta}=\mel{\psi_\alpha}{\qop{O}}{\psi_\beta}$ of a few-body observable $\qop{O}$ in the Hamiltonian eigenstates $\ket{\psi_\alpha}$ take the form
\begin{equation}\label{eq:ETH_ansatz}
    O_{\alpha\beta} = O_{\mathrm{mc}}(E_\alpha)\delta_{\alpha \beta} + \Omega^{-1/2}(\bar{E}_{\alpha\beta}) f^{}_O(\bar{E}_{\alpha\beta}, \omega^{}_{\alpha\beta}) \, R_{\alpha \beta}\,.
\end{equation}
For a given pair of eigenenergies $E_\alpha$ and $E_\beta$, $\bar{E}_{\alpha\beta} \equiv (E_\alpha + E_\beta) / 2$ and $\omega^{}_{\alpha\beta} \equiv E_\alpha - E_\beta$ (we set $\hbar=1$). $O_{\mathrm{mc}}(E_\alpha)$ is the microcanonical expectation value of $\qop{O}$ at energy $E_\alpha$ and $f^{}_O(\bar{E}_{\alpha\beta}, \omega^{}_{\alpha\beta})$ is a smooth function that controls the variance of off-diagonal matrix elements. $\Omega(\bar{E}_{\alpha\beta}) \propto \exp[S(\bar{E}_{\alpha\beta})]$ is the many-body density of states and $S(\bar{E}_{\alpha\beta})$ is the thermodynamic entropy. For Hamiltonians that are time-reversal invariant, on which we focus here, $R_{\alpha \beta}$ is a random variable with zero mean and unit variance (variance 2) for $\alpha \neq \beta$ ($\alpha=\beta$).

The ETH spectral function $\abs{f^{}_O(\bar{E}, \omega)}^2$ is obtained by averaging $\big|f^{}_O(\bar{E}_{\alpha\beta}, \omega^{}_{\alpha\beta})\big|^2$ in suitable windows in which pairs of energy eigenstates have $\bar{E}_{\alpha\beta}\approx \bar E$ and $\omega^{}_{\alpha\beta}\approx \omega$. $\abs{f^{}_O(\bar{E}, \omega)}^2$ is, in general, expected to exhibit a plateau for $\omega^{}_{\mathrm{H}} < \omega \lesssim \omega^{}_{\mathrm{ETH}}$ [see Fig.~\ref{fig:schematic}(c)], where $\omega^{}_{\mathrm{H}}=2\pi/t^{}_{\mathrm{H}}$ is the level spacing at energy $\bar{E}$~\cite{DAlessio2016, LeBlond2020, Richter2020, Patil2026}. In that regime, the statistics of the matrix elements are anticipated to exhibit RMT behavior. Furthermore, $\omega^{}_{\mathrm{ETH}}$ is expected to inherit its $L$-dependence from the smallest transport frequency $\omega^{\mathrm{ch}^{\!\star}}_{\mathrm{tr}}=2\pi/t^{\mathrm{ch}^{\!\star}}_{\mathrm{tr}}$ in diffusive systems~\cite{DAlessio2016, Luitz2016, LeBlond2020, Schonle2021, Sels2021, Capizzi2025, Ceven2026}, as well as in sub- and superdiffusive systems. We note that other criteria have been used to define RMT behavior~\cite{Dymarsky2022, Wang2022}, and that RMT behavior has also been characterized using out-of-time ordered correlators~\cite{Brenes2021}.

The goal of this work is to directly compare the frequencies $\omega^{}_{\mathrm{SFF}}$, $\omega^{\mathrm{E}}_{\mathrm{tr}}$, $\omega^{\mathrm{S}}_{\mathrm{tr}}$, and $\omega^{}_{\mathrm{ETH}}$, (where the superscripts E and S denote the transport channels corresponding to energy and spin transport, respectively). Using exact diagonalization, we consider disordered spin-$\tfrac{1}{2}$ XX ladders and clean spin-$\tfrac{1}{2}$ XXZ ladders in parameter regimes in which observables are well described by the ETH. In the models discussed here, there are two hydrodynamic channels corresponding to the conservation of energy and total magnetization. Our focus is on spectral functions $\abs{f^{}_O(E_\infty, \omega)}^2$ for the associated spin- and energy-current operators. We compute these at $\bar E=E_\infty$, the energy corresponding to infinite temperature. Since there is no established procedure for determining $\omega^{}_{\mathrm{ETH}}$ from $\abs{f^{}_O(E_\infty, \omega)}^2$~\cite{DAlessio2016, Vidmar2021, Patil2026}, we will compute the transport frequencies and $\omega^{}_{\mathrm{SFF}}$ and discuss their location in the spectral function.

Spin-$\tfrac12$ XX ladders are relevant in the context of recent quantum-simulator experiments~\cite{Ronzheimer2013, Wienand2024, Hur2025, Lunkin2026} and their transport properties have been studied in the clean~\cite{Steinigeweg2014C, Kloss2018, Rakovszky2022, Fitzner2026} and disordered~\cite{Ceven2026} cases. For disordered XX ladders, the energy and spin-transport times were found to be shorter than $t^{}_{\mathrm{SFF}}$ \cite{Ceven2026}. Moreover, energy diffusion was found to be the slowest channel~\cite{Ceven2026}. Based on these results, we anticipate the same hierarchy in the models studied here, i.e.,~$\omega^{}_{\mathrm{SFF}}< \omega^{\mathrm{E}}_{\mathrm{tr}}< \omega^{\mathrm{S}}_{\mathrm{tr}}$.

For clean systems, such as our ladders, numerical calculations of the SFF can suffer from large fluctuations since the SFF is not self-averaging~\cite{Prange1997, Braun2015}. As a promising byproduct of our study, we find that averaging the SFFs calculated in different symmetry sectors of the Hamiltonian suppresses fluctuations, allowing us to clearly resolve the RMT ramp in the SFF (for different strategies, see Refs.~\cite{Matsoukas-Roubeas2023, Charamis2026}). We also stress that, in this work, we label all frequencies and time scales according to how they are determined, i.e.,~via the spectral form factor, transport, or the ETH spectral function. In the literature, the term Thouless frequency is used interchangeably for these three frequencies, so we do not use it here to avoid ambiguities.

This paper is organized as follows. In Sec.~\ref{sec:model}, we introduce the clean and disordered models considered in this work and motivate the parameters used in our study. We introduce the observables and quantities of interest, and discuss how they are computed, in Sec.~\ref{sec:observables}. A detailed explanation of some of the calculations involved is then provided in the Appendices. In Sec.~\ref{sec:results}, we report and discuss our results. A summary of our findings and an outlook are provided in Sec.~\ref{sec:conclusions}.

\section{Model}\label{sec:model}

We consider spin-$\tfrac{1}{2}$ XXZ ladders with Hamiltonian:
\begin{equation}\label{eq:XX_ladder_ham}
\qop{H} = \qop{H}_\parallel + \qop{H}_\perp + \qop{H}_{\mathrm{dis}} + \qop{H}_{\Delta} \,.
\end{equation}

The first two operators, $\qop{H}_\parallel$ and $\qop{H}_\perp$, describe spin-flip terms along the legs and the rungs, respectively:
\begin{align}
    \qop{H}_\parallel &= \frac{J_\parallel}{2} \sum_{\ell=1}^L \sum_{i=1}^2 \qty(\qop{S}^+_{\ell, i} \qop{S}^-_{\ell+1, i} + \qop{S}^-_{\ell, i} \qop{S}^+_{\ell+1, i})\,,\\
    \qop{H}_\perp &= \frac{J_\perp}{2} \sum_{\ell=1}^L \qty(\qop{S}^+_{\ell, 1} \qop{S}^-_{\ell, 2} + \qop{S}^-_{\ell, 1} \qop{S}^+_{\ell, 2})\,,
\end{align}
where $J_\parallel$ and $J_\perp$ are the intra-leg and rung coupling strengths, respectively, and $L$ is the number of rungs, i.e.,~the total number of spins is $2L$. $\qop{S}^\pm_{\ell, i} = \qop{S}^x_{\ell, i} \pm i \qop{S}^y_{\ell, i}$ are the raising and lowering operators acting at site $(\ell, i)$. The third operator $\qop{H}_{\mathrm{dis}}$ describes quenched disorder:
\begin{equation}\label{eq:XX_ladder_ham_disorder_part}
    \qop{H}_{\mathrm{dis}} = \sum_{\ell=1}^L \sum_{i=1}^2 w^{}_{\ell, i} \qop{S}^z_{\ell, i}\,,
\end{equation}
where $\qop S^z_{\ell,i}$ is the $z$-component of a spin-$\tfrac{1}{2}$ operator. The random local magnetic-field strengths $w^{}_{\ell, i}$ are sampled independently from a uniform distribution on $[-W, W]$, with $W \geq 0$ controlling the disorder strength. The final term $\qop{H}_{\Delta}$ introduces an Ising interaction (with coupling strength $\Delta$) along the legs and reads:
\begin{equation}
    \qop{H}_\Delta =  \,\Delta \sum_{\ell=1}^L \sum_{i=1}^2 \qop{S}^z_{\ell, i} \qop{S}^z_{\ell+1, i} \,.\label{eq:Delta}
\end{equation}
We include $\qop H_\Delta$ only in the clean case to obtain a model whose eigenstates satisfy the ETH, since clean spin-$\tfrac12$ XX ladders, described by $\qop H_{\rm XX}=\qop H_\parallel+\qop H_\perp$, exhibit Hilbert-space fragmentation~\cite{Znidaric2013A, Znidaric2013B}.

We impose periodic boundary conditions along the legs, $\qop{\vec S}_{L+1,i}\equiv \qop{\vec S}_{1,i}$. For both the clean and disordered cases, we restrict the analysis to the zero-magnetization sector ($S^z=0$) unless stated otherwise. In the clean case, we also resolve the translational symmetry along the legs labeled by total quasimomentum $k^{}_x$, the reflection symmetry along the rungs labeled by $p^{\mathrm{r}}_y$, and the spin-$z$ parity symmetry labeled by $p^{\mathrm{S}}_z$. Disorder, see Eq.~\eqref{eq:XX_ladder_ham_disorder_part}, breaks these symmetries except for the conservation of total magnetization. In our calculations on disordered ladders, we average over 500 disorder realizations unless stated otherwise.

Obtaining results for the spectral functions that exhibit reduced finite-size effects at low frequencies for the ladder sizes considered requires identifying the parameter regime of our Hamiltonian for which the statistical properties of the eigenenergies and eigenstates are closest to the RMT predictions. The low-frequency behavior close to integrable and localized regimes can be starkly different depending on the system size considered~\cite{Pandey2020, Sels2021, LeBlond2021, Bulchandani2022, Orlov2023, Kim2024, Kim2025, Abdelshafy2025}. Various indicators, e.g.,~based on eigenenergies~\cite{Oganesyan2007, Santos2010, Mondaini2017, Kliczkowski2023}, eigenstate components~\cite{Santos2010, Beugeling2018, Haque2022, Kliczkowski2023}, matrix elements of observables~\cite{Mondaini2017, LeBlond2019, LeBlond2020, Fogarty2021}, and eigenstate entanglement entropies~\cite{Kliczkowski2023, Rodriguez-Nieva2024}, have been used to quantify the agreement with RMT predictions.

We focus on an indicator based on the eigenstate components. For a symmetry sector $s$ with quantum numbers $\{q\}$, we compute the ratio
\begin{equation}\label{eq:gaussianity}
    \Gamma_s = \frac{\overline{\big|x_n^{(\alpha)}\big|^2}}{\qty(\,\overline{\big|x_n^{(\alpha)}\big|}\,)^{\!2}}\,,
\end{equation}
where $x_n^{(\alpha)}$ is the real part of the expansion coefficient $C_n^{(\alpha)}$ of the eigenstate $\lvert \psi_\alpha\rangle$ in the basis $\{\lvert {\{ q \}}_n\rangle\}$, $\lvert \psi_\alpha\rangle = \sum_n C_n^{(\alpha)} \, \lvert {\{ q \}}_n\rangle$. The average is taken over $\lfloor\min[100, 0.1 \times \mathcal{D}({\{ q \}})]\rfloor$ mid-spectrum energy eigenstates and over all the basis states.

For the disordered case, we average $\Gamma_s$ over 500 disorder realizations, whereas for the clean case, we average over symmetry sectors weighted by the corresponding symmetry-sector dimensions. We call the result of the average $\Gamma$. For Gaussian-distributed coefficients, as expected for eigenstates of random matrices, $\Gamma=\pi/2$. The deviation of $\Gamma$ from $\pi/2$ signals a departure from RMT behavior. We also studied the mean adjacent-gap ratios~\cite{Oganesyan2007, Atas2013} and the mean eigenstate entanglement entropies, and found $\Gamma$ to be a better indicator in our ladders.

\begin{figure}
\includegraphics[width=\linewidth]{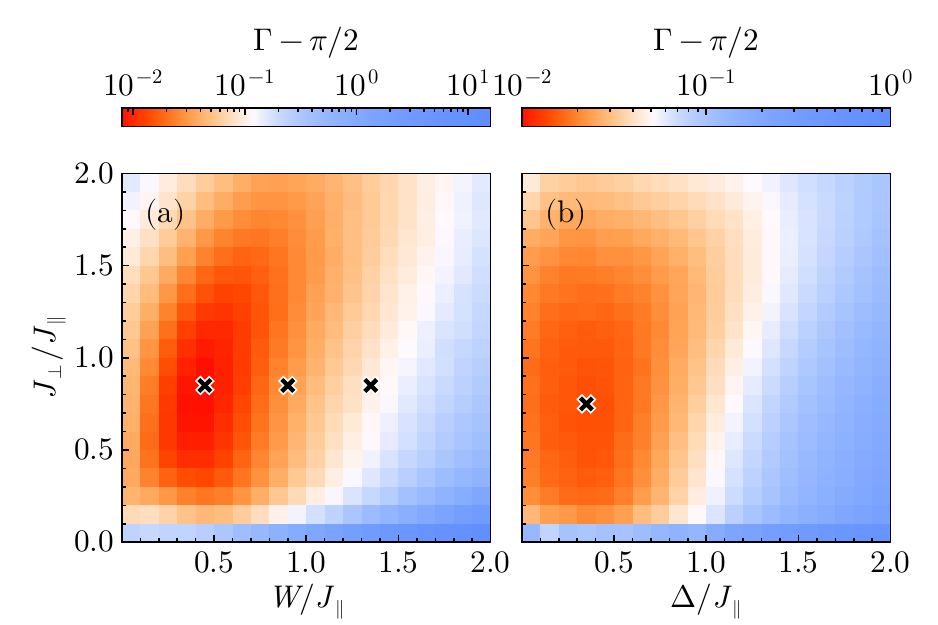}
\vspace{-0.5cm}
\caption{$\Gamma - \pi/2$ (color scale) for (a) a disordered spin-$\tfrac{1}{2}$ XX ladder as a function of the disorder strength $W/J_\parallel$ and the coupling ratio $J_\perp/J_\parallel$ for $L=8$ and (b) a clean spin-$\tfrac{1}{2}$ XXZ ladder as a function of the intra-leg anisotropy strength $\Delta / J_\parallel$ and the coupling ratio $J_\perp/ J_\parallel$ for $L=11$. The crosses show the parameters considered in this work.}
\label{fig:mcr}
\end{figure}

In Fig.~\ref{fig:mcr}, we plot $\Gamma - \pi/2$ as a function of the disorder strength $W/J_\parallel$ and the coupling ratio $J_\perp/J_\parallel$ for a disordered ladder with $L=8$ [see Fig.~\ref{fig:mcr}(a)] and as a function of the intra-leg anisotropy strength $\Delta/J_\parallel$ and the coupling ratio $J_\perp/J_\parallel$ for a clean ladder with $L=11$ [see Fig.~\ref{fig:mcr}(b)]. Motivated by these results, for our calculations of disordered ladders, we fix $J_\perp/J_\parallel = 0.85$ and take $W/J_\parallel = 0.45$, where $\Gamma$ is closest to $\pi/2$. We also consider $W/J_\parallel= 0.9$ and $1.35$ to study the effect of increasing the disorder strength on the spectral functions, and on $\omega^{}_{\mathrm{SFF}}$, $\omega^{\mathrm{E}}_{\mathrm{tr}}$, and $\omega^{\mathrm{S}}_{\mathrm{tr}}$. For the clean case, we take $J_\perp/J_\parallel = 0.75$ and $\Delta/J_\parallel = 0.35$. These parameters are marked by crosses in Fig.~\ref{fig:mcr}. Similar results are obtained for the other ladder sizes considered in this work.

\section{Observables}\label{sec:observables}

\subsection{Spectral form factor}\label{sec:rmt_beha_sff}

To determine $\omega^{}_{\mathrm{SFF}}$ from the time dependence of the SFF, we evaluate~\cite{Berry1985, Haake1991, Guhr1998, Stoeckmann1999, Mueller2004, Mueller2005, Kos2018, Chan2018, Sierant2025},
\begin{equation}
    K(t) = \abs{\sum_\alpha \exp(-i E_\alpha t )}^2\,.
\end{equation}
In a generic quantum many-body system, $K(t)$ approaches a characteristic RMT ramp at times $t\gtrsim t^{}_{\mathrm{SFF}}$~\cite{Bertini2018}.

For the numerical evaluation of the SFFs, we diagonalize the Hamiltonian and use unfolding. The extraction of $t^{}_{\mathrm{SFF}}$ involves two parameters: the width $\eta$ of the Gaussian filter and the threshold $\epsilon_{\Delta K}$ used to determine when $K(t)$ reaches the RMT prediction. These parameters, together with the limited accessible ladder sizes $L$, yield the reported uncertainties in the extraction of $t^{}_{\mathrm{SFF}}$. For cases with disorder, we average over 3500 disorder realizations, while in the clean case, we calculate the SFF within each symmetry sector of the Hamiltonian and then average the results. Finally, we calculate $t^{}_\mathrm{H}$ using the mean level spacing in the central $10\%$ of the energy spectrum, where the density of states varies relatively weakly. Note that for a smaller percentage of states than 10\%, the Heisenberg times are slightly larger by a few percent. See Appendix~\ref{sec:sff} for a detailed discussion of the calculations.

\subsection{Current operators}\label{sec:currentoperators}

The total spin and energy currents are obtained from the continuity equations for the corresponding conserved quantities. For a given transport channel, we define the Hilbert-Schmidt-scaled total current as~\cite{Mierzejewski2020, Lydzba2024}
\begin{equation}\label{eq:tot_current}
    \qop{J}^{\mathrm{ch}} = \frac{1}{\sqrt{2L}} \sum_{\ell=1}^L \qop{j}_\ell^{\mathrm{ch}}\,,
\end{equation}
where $2L$ is the number of lattice sites in our ladders.

\paragraph{Spin current} To find the local spin-current operator $\qop{j}^{\mathrm{S}}_\ell$, we decompose the total magnetization into a sum of local magnetizations as $\qop{Q}^{\mathrm{S}} = \qop{S}^z = \sum_\ell \qop{q}_\ell^{\mathrm{S}}$, where the local magnetization at the $\ell$th rung is defined as $\qop{q}_\ell^{\mathrm{S}} \coloneqq \sum_i \qop{S}^z_{\ell, i} = \qop{S}^z_{\ell, 1} + \qop{S}^z_{\ell, 2}$. Using the lattice continuity equation, we find
\begin{equation}
    \qop{j}^{\mathrm{S}}_\ell = J_\parallel \sum_{i=1}^2 \qop{j}^{}_{(\ell, \ell+1), (i, i)}\,,
\end{equation}
where the local current $\qop{j}^{}_{(\ell, \ell^\prime), (i, i^\prime)}$ is defined as
\begin{equation}
    \qop{j}^{}_{(\ell, \ell^\prime), (i, i^\prime)} \coloneqq \frac{i}{2} \left( \qop{S}_{\ell, i}^+ \qop{S}_{\ell^\prime, i^\prime}^- - \qop{S}_{\ell, i}^- \qop{S}_{\ell^\prime, i^\prime}^+ \right)\,.
\end{equation}

\paragraph{Energy current} Likewise, to find the local energy-current operator $\qop{j}_\ell^\mathrm{E}$, we write $\qop{Q}^{\mathrm{E}} = \qop{H} = \sum_\ell \qop{q}^{\mathrm{E}}_\ell$. Using the definition of $\qop{q}_\ell^{\mathrm{E}}$ adopted in Ref.~\cite{Ceven2026, Steinigeweg2014C}, which we extend to include the anisotropy term, we find
\begin{equation}\label{eq:loc_energy_current}
    \qop{j}_\ell^\mathrm{E} = \qop{j}_{\parallel,\ell}^\mathrm{E} + \qop{j}_{\perp,\ell}^\mathrm{E} + \qop{j}_{\Delta,\ell}^\mathrm{E}\,,
\end{equation}
where the intra-leg $\qop{j}^{\mathrm{E}}_{\parallel, \ell}$, rung $\qop{j}^{\mathrm{E}}_{\perp, \ell}$, and anisotropy $\qop{j}_{\Delta,\ell}^\mathrm{E}$ contributions are given by
\begin{widetext}
\begin{align}
    \qop{j}_{\parallel,\ell}^\mathrm{E} &= J^2_\parallel \sum_{i=1}^2 \qop{j}^{}_{(\ell+1, \ell-1), (i, i)} \qop{S}^z_{\ell, i} + J_\parallel \sum_{i = 1}^2 \frac{w^{}_{\ell, i} + w^{}_{\ell+1, i}}{2} \qop{j}^{}_{(\ell, \ell+1), (i, i)}\,,\\
    \qop{j}_{\perp,\ell}^\mathrm{E} &= \frac{J_\parallel J_\perp}{2} \sum_{i=1}^2 \qop{j}^{}_{(\ell+1, \ell), (i, 3-i)} \qop{S}^z_{\ell, i} - \frac{J_\parallel J_\perp}{2} \sum_{i=1}^2 \qop{j}^{}_{(\ell-1, \ell), (i, 3-i)} \qop{S}^z_{\ell, i}\,,\\
    \qop{j}_{\Delta,\ell}^\mathrm{E} &= J_\parallel\Delta\sum_{i=1}^2 \left( \qop{j}^{}_{(\ell-1, \ell), (i, i)} \qop{S}^z_{\ell+1, i} + \qop{j}^{}_{(\ell, \ell+1), (i, i)} \qop{S}^z_{\ell-1, i} \right) + \frac{J_\perp \Delta}{2} \sum_{i=1}^2 \qop{j}^{}_{(\ell, \ell), (i, 3-i)} \left( \qop{S}^z_{\ell-1, i} - \qop{S}^z_{\ell+1, i}\right)\,.
\end{align}
\end{widetext}

\subsection{Spectral functions}\label{sec:off-diag}

The central objects of interest are the spectral functions $\abs{f^{}_{O}(\bar{E}, \omega)}^2$ for the total energy-current ($\qop{O} = \qop{J}^\mathrm{E}$) and spin-current ($\qop{O} = \qop{J}^\mathrm{S}$) operators. These quantities can either be computed using the connected autocorrelation functions via the fluctuation-dissipation theorem~\cite{DAlessio2016, Patil2026A} or directly from the off-diagonal matrix elements at fixed $\bar{E}$ and $\omega$. We use the latter approach in this work. We consider $\bar{E}= E_\infty$, the energy corresponding to infinite temperature. For a symmetry sector of dimension $\mathcal{D}$, $E_\infty$ is obtained from $E_\infty \equiv \langle \qop{H} \rangle=(\Tr\qop{H})/\mathcal{D}$.

For an observable $\qop O$, we can write~\cite{Patil2026A}
\begin{equation}\label{eq:off-diag_spec_func}
    \abs{f^{}_O(E_\infty, \omega)}^2 \approx \Omega(E_\infty,\Delta E) \, \overline{\abs{O_{\alpha \beta}}^2} \, ,
\end{equation}
where $\overline{\qty(\,\cdots)}$ is a centered moving average with logarithmic frequency windows (see Appendix~\ref{sec:cal_f_O}). Since we average over a narrow energy window of half width $\Delta E \sqrt{2L}$ around $E_\infty $ in each given symmetry sector, $\Omega{(E_\infty, \Delta E)} $ is the density of states in the chosen energy window. For the clean model, we carry out a weighted ensemble average over symmetry sectors within the zero-magnetization sector. Since the $k^{}_x=0$ and $k^{}_x =\pi$ total quasimomentum sectors have an additional parity symmetry, which results in stronger finite-size effects, we exclude those sectors when computing the weighted average~\cite{LeBlond2020}.

\subsection{Diffusion constants}\label{sec:transport_freqs}

For systems with diffusive spin and energy transport, the transport frequencies $\omega_{\mathrm{tr}}^{\mathrm{E}}$ and $\omega_{\mathrm{tr}}^{\mathrm{S}}$ are obtained from the diffusion constants $D^{\mathrm{E}}$ and $D^{\mathrm{S}}$, respectively. In the Kubo formalism~\cite{Kubo1957, Bertini2021}, the diffusion constants can be extracted from a computation of the autocorrelation function of the total current operator ($\mathrm{ch}=\mathrm{E},\mathrm{S}$):
\begin{equation}\label{eq:autocorrelation_C_J}
    C^{\mathrm{ch}}{(t)} = \Re{\expval{\qop{J}^{\mathrm{ch}}{(t)} \, \qop{J}^{\mathrm{ch}}{(0)} }}\,,
\end{equation}
where $\expval{\cdots}$ stands for the thermal expectation at infinite temperature. The diffusion constant is then obtained by integrating $C^{\mathrm{ch}}(t)$ over $t\in [0,\infty)$ and dividing by the leading coefficient in the high-temperature expansion of the static susceptibility for each transport channel, $\chi^\mathrm{ch}\equiv\tfrac{1}{2L}[\langle (\qop{Q}^{\mathrm{ch}})^2\rangle-\langle\qop{Q}^{\mathrm{ch}}\rangle^2]$:
\begin{equation}\label{eq:diff_const}
    D^{\mathrm{ch}} = \frac{1}{\chi^\mathrm{ch}} \int_0^\infty \dd{t} \, C^{\mathrm{ch}}(t)\,.
\end{equation}
This expression assumes that the thermodynamic limit is taken before the infinite-time limit. In practice, since our numerical simulations are inherently restricted to finite system sizes, both the accessible time scales and possible finite-size dependencies impose limitations. As is standard practice for finite systems \cite{Bertini2021}, we compute the time-dependent version, $D^{\mathrm{ch}}(t)$, by replacing the upper limit of the integration in Eq.~\eqref{eq:diff_const} with a finite time $t$. To estimate $D^{\mathrm{ch}}$ from $D^{\mathrm{ch}}(t)$, we average the latter over a late-time interval \cite{Steinigeweg2009, Steinigeweg2014C}. To compute the disorder-averaged diffusion constant, we average $C^{\mathrm{ch}}(t)$ over disorder realizations.

$\chi^\mathrm{ch}$ is straightforward to calculate analytically in the thermodynamic and infinite-temperature limits. For the spin channel of clean and disordered spin-$\tfrac12$ XX and XXZ ladders
\begin{equation}\label{eq:spin_suscep}
    \chi_{\mathrm{XX}}^\mathrm{S} = \chi_{\mathrm{XXZ}}^\mathrm{S} = \frac{1}{2L}\qty[\ev{\qty(\qop{S}^{z})^{\!2}}-\ev{\qop{S}^{z}}^{\!2}] = \frac{1}{4}\,.
\end{equation}
For the energy channel of disordered spin-$\tfrac{1}{2}$ XX ladders~\cite{Ceven2026},
\begin{equation}\label{eq:disordered_energy_suscep}
    \chi_{\mathrm{XX}}^\mathrm{E} = \frac{1}{2L}\qty[\ev{\qop{H}^2}-\ev{\qop{H}}^{\!2}] = \frac{3( 2J^2_\parallel + J^2_\perp ) + 4 W^2}{48}\,,
\end{equation}
whereas, for clean spin-$\tfrac{1}{2}$ XXZ ladders,
\begin{equation}\label{eq:clean_energy_suscep}
    \chi_{\mathrm{XXZ}}^\mathrm{E} = \frac{2J^2_\parallel + J^2_\perp + \Delta^2}{16}\,.
\end{equation}
In the canonical ensemble, for $S^z=0$, $\chi_{\mathrm{XX}}^\mathrm{ch}$ and $\chi_{\mathrm{XXZ}}^\mathrm{ch}$ have finite-size corrections. They are not significant in the scale of our plots. 

In our numerical calculations of the diffusion constants, we use the grand-canonical ensemble (i.e.,~average over all $S^z$ sectors) and employ the dynamical quantum-typicality (DQT) approach~\cite{Hams2000, Bartsch2009, Elsayed2013, Steinigeweg2014A, Steinigeweg2014B, Steinigeweg2015, Steinigeweg2016} combined with Krylov-space time evolution~\cite{Manmana2005}. These results are complemented by results from the canonical ensemble ($S^z=0$ sector only), allowing us to estimate finite-size uncertainties. See Appendix~\ref{sec:diff_const} for a detailed discussion of the calculations.

\section{Results}\label{sec:results}

\subsection{Disordered spin-\texorpdfstring{$\tfrac{1}{2}$}{1/2} XX ladders}\label{sec:resultsdisorder}

\begin{figure}
\includegraphics[width=\linewidth]{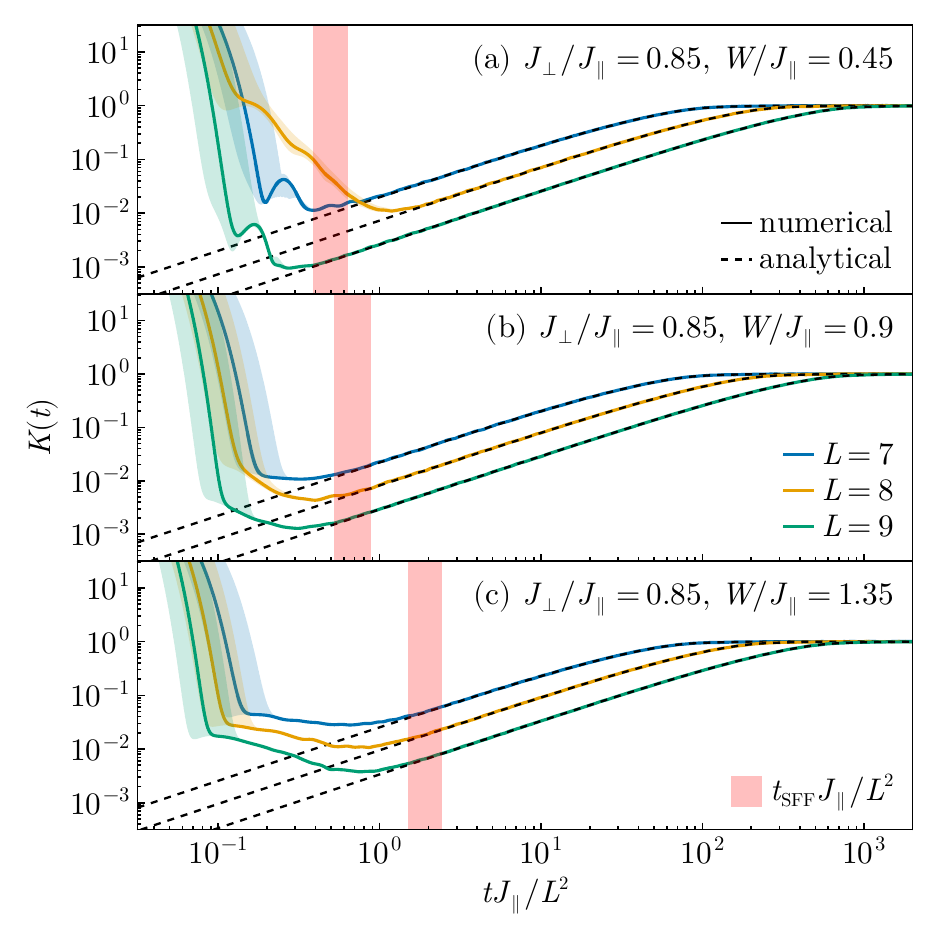}
\vspace{-0.7cm}
\caption{SFF $K(t)$ of disordered spin-$\tfrac{1}{2}$ XX ladders versus normalized time for $J_\perp/J_\parallel = 0.85$ and disorder strengths: (a) $W/J_\parallel = 0.45$, (b) $W/J_\parallel = 0.9$, and (c) $W/J_\parallel = 1.35$. The solid lines show the numerical results for the SFFs obtained for $\eta = 0.3$ and the dashed lines show the analytical GOE prediction $K_\mathrm{GOE}(t)$. All numerical SFFs are averaged over $M=3500$ disorder realizations. The shaded bands around the SFF results, mainly visible in the nonuniversal short-time regime, show the range of SFF curves obtained for $\eta \in [0.2, 0.4]$. The vertical shaded regions show the range of SFF times $t^{}_\mathrm{SFF}$, extracted from the largest-accessible (linear) ladder size $L=9$ using $\eta \in [0.2, 0.4]$, and $\epsilon^{}_{\Delta K} \in [0.005, 0.025]$, further widened to take into account the uncertainty in $t^{}_\mathrm{H}$. See Appendix~\ref{sec:sff} for details.}
\label{fig:sff_vs_t_disordered}
\end{figure}

\begin{figure*}
\includegraphics[width=\linewidth]{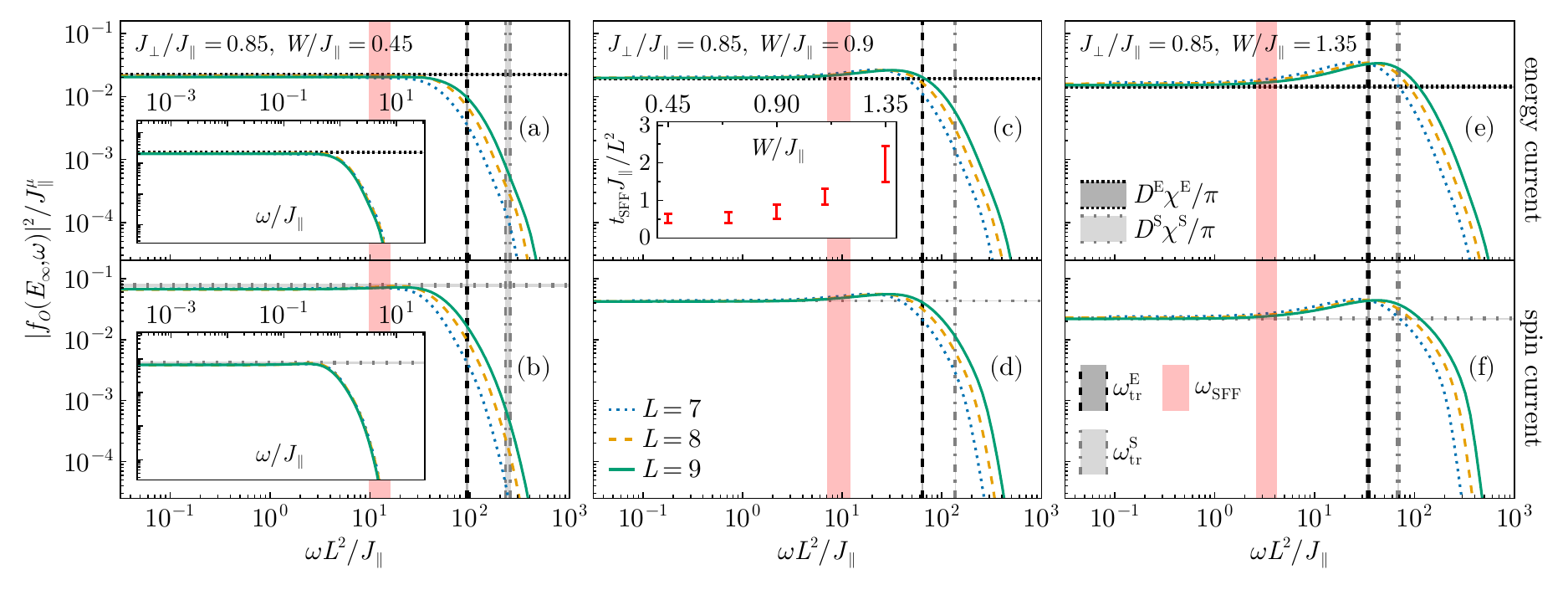}
\vspace{-0.7cm}
\caption{Spectral function of disordered spin-$\tfrac{1}{2}$ XX ladders $\abs{f^{}_{O}(E_\infty, \omega)}^2$ versus $\omega L^2$ for (a, c, e) the energy current $\qop{O}=\qop{J}^{\mathrm{E}}$ and (b, d, f) the spin current $\qop{O}=\qop{J}^{\mathrm{S}}$, for $J_\perp/J_\parallel = 0.85$ and disorder strengths (a, b) $W/J_\parallel = 0.45$, (c, d) $W/J_\parallel = 0.9$, and (e, f) $W/J_\parallel = 1.35$. The spectral functions are expressed in units of $J_\parallel^\mu$, $\mu =1$ for the spin and $\mu=3$ for the energy current. Results are shown for ladders with $L = 7, 8, 9$ rungs. The vertical shaded regions show the range of SFF frequencies $\omega^{}_{\mathrm{SFF}}=2\pi/t^{}_\mathrm{SFF}$ corresponding to the range of SFF times from Fig.~\ref{fig:sff_vs_t_disordered}, while dashed and dot-dashed lines indicate the transport frequencies $\omega_{\mathrm{tr}}^\mathrm{E}$ and $\omega_{\mathrm{tr}}^\mathrm{S}$, respectively. The horizontal lines show the result from Eq.~\eqref{eq:fzero}, i.e.,~$D^{\mathrm{ch}} \chi^{\mathrm{ch}}/ \pi$, where $D^\mathrm{ch}$ is independently obtained from the Kubo formula. The narrow shaded bands around the vertical and horizontal lines show the uncertainties in the determination of these quantities, which in some cases are within the linewidth. The insets in (a, b) show the spectral functions vs $\omega$ (without $L^2$ rescaling). The inset in (c) shows $t^{}_{\mathrm{SFF}}$ divided by $L^2$ plotted as a function of $W$.}
\label{fig:off-diag_ETH_vs_omega_L_squared}
\end{figure*}

The disorder strengths chosen for the discussion in this section, see Fig.~\ref{fig:mcr}, are much smaller than those required to enter the crossover into the finite-size many-body localized regime~\cite{Ceven2026}. In Fig.~\ref{fig:sff_vs_t_disordered}, we show our results for the SFF, obtained as explained in detail in Appendix~\ref{sec:sff}. For all disorder strengths and ladder sizes shown, one can see the expected nonuniversal regime at short times, followed by the universal RMT ramp, and then the plateau beyond the Heisenberg time $t^{}_\mathrm{H}$. $t^{}_\mathrm{SFF}$ marks the onset of the RMT ramp and our results are consistent with the expected scaling $t^{}_\mathrm{SFF}\propto L^2$. Note that the time axes are rescaled by $L^2$. We extract $t^{}_\mathrm{SFF}$ from the largest available (linear) system size, $L=9$, and report as shaded bands the results obtained for $\eta \in [0.2, 0.4]$ and $\epsilon^{}_{\Delta K} \in [0.005, 0.025]$, further widened to take into account the uncertainty in $t^{}_\mathrm{H}$ associated with the energy window used to compute it (see Appendix~\ref{sec:sff}). The bands therefore reflect uncertainty in the extraction procedure itself, not a statistical error. 

Our results in Fig.~\ref{fig:sff_vs_t_disordered} show that, for the disorder strengths considered, the (logarithmic scale) uncertainty in the extraction of $t^{}_\mathrm{SFF}$ does not depend much on the disorder strength. The SFF for $L=8$ in the ladders with the weakest disorder [see Fig.~\ref{fig:sff_vs_t_disordered}(a)] deviates from the rest of the results. We attribute this to the emergence of an increasing number of mid-spectrum near-degeneracies as the disorder strength decreases for even values of $L$. The latter is a remnant of the Hilbert-space fragmentation~\cite{Znidaric2013A} in the clean limit of our spin-$\tfrac12$ ladders. We also find that, for the disorder strengths considered, the band of values of $t^{}_\mathrm{SFF}$ is consistent with a monotonic increase with $W$; see the inset of Fig.~\ref{fig:off-diag_ETH_vs_omega_L_squared}(c) [see also Fig.~\ref{fig:sff_vs_t_disordered}(c)]. This, in turn, is consistent with the expected increase of the transport times resulting from the decrease of the diffusion constants as the disorder strength increases.

In the insets in Figs.~\ref{fig:off-diag_ETH_vs_omega_L_squared}(a) and~\ref{fig:off-diag_ETH_vs_omega_L_squared}(b), we plot $\abs{f^{}_{O}(E_\infty,\omega)}^2$ as a function of $\omega$ for the energy and spin current, respectively. The emergence of a low-frequency plateau is evident in all system sizes, with no discernible finite-size effects in the plateau height. A straightforward calculation~\cite{Schonle2021, Marcin-private} shows that the value of the plateau is 
\begin{equation}
\lim_{\omega \to 0} \abs{f^{}_{O=J^{\mathrm{ch}}}(E_\infty,\omega)}^2 = \frac{D^{\mathrm{ch}} \chi^{\mathrm{ch}} }{ \pi}\,. \label{eq:fzero}
\end{equation}
Thus, the value of the spectral function in the plateau is proportional to the respective diffusion constant. By using our independent calculation of the diffusion constants from the Kubo formula (horizontal dotted lines in the insets), we have verified Eq.~\eqref{eq:fzero} within the accuracy of the finite-size methods. In our ladders, the decay of the spectral functions at high frequencies is well described by a stretched exponential. Exponential and Gaussian decays have been reported in the literature for integrable and nonintegrable chains~\cite{LeBlond2019, Jansen2019, LeBlond2020, Zhang2022}.

To connect the extent of the plateaus of the spectral function to the SFF and transport frequencies, in the main panels in Fig.~\ref{fig:off-diag_ETH_vs_omega_L_squared}, we rescale the frequency axes by $L^2$. In Figs.~\ref{fig:off-diag_ETH_vs_omega_L_squared}(a,c,e), we show the spectral functions for the energy current, and in Figs.~\ref{fig:off-diag_ETH_vs_omega_L_squared}(b,d,f) for the spin current, as the disorder strength increases. In all cases, the plateaus are seen at values of the spectral function that are consistent with the predictions of Eq.~\eqref{eq:fzero}. 

In Fig.~\ref{fig:off-diag_ETH_vs_omega_L_squared}, the vertical red shaded regions indicate the values of $\omega^{}_{\mathrm{SFF}}=2\pi/t^{}_\mathrm{SFF}$ corresponding to the values of $t^{}_\mathrm{SFF}$ in Fig.~\ref{fig:sff_vs_t_disordered}. For all disorder strengths shown in Fig.~\ref{fig:off-diag_ETH_vs_omega_L_squared}, those regions coincide with the end of the plateaus on the accessible system sizes. Therefore, we find that $\omega^{}_{\mathrm{SFF}}$ is consistent with $\omega^{}_{\mathrm{ETH}}$. We note that since the plateau values depend on the diffusion constants, which themselves must be determined numerically, there is no exact benchmark for the spectral function analogous to the analytical RMT prediction for the SFF ramp. Therefore, given that the spectral functions are rather featureless in the region in which the plateau ends, it is not obvious how one could extract $\omega^{}_{\mathrm{ETH}}$ from the spectral functions alone. Also, even after rescaling the frequency axis with $L^2$, the width of the plateaus exhibits a weak increase with system size.

In Fig.~\ref{fig:off-diag_ETH_vs_omega_L_squared}, we also show the transport frequencies for energy (vertical dashed lines) and spin (vertical dot-dashed lines). Those values (lines) and their uncertainties (shaded bands around the lines) are obtained in Appendix~\ref{sec:diff_const}. Figure~\ref{fig:off-diag_ETH_vs_omega_L_squared} shows that the transport frequencies are larger than $\omega^{}_{\mathrm{SFF}}$, as expected from the results in Ref.~\cite{Ceven2026}, and lie in the nonuniversal regime of the spectral functions. Note that, as anticipated, the smallest transport frequency (corresponding to the largest transport time) is the one related to energy diffusion~\cite{Ceven2026}. Hence, for the disordered ladders considered in this work and for the accessible system sizes:
\begin{align}\label{eq:relationsbetweenomegas}
\omega^{}_{\mathrm{SFF}} \approx \omega^{}_{\mathrm{ETH}} < \omega_{\mathrm{tr}}^{\mathrm{E}} < \omega_{\mathrm{tr}}^{\mathrm{S}}\,.
\end{align}

\subsection{Clean spin-\texorpdfstring{$\tfrac{1}{2}$}{1/2} XXZ ladders} 

The conclusions in Sec.~\ref{sec:resultsdisorder} are robust against changes of the disorder strength in the diffusive regime of our ladders. In this section, we probe their robustness in the clean case ($W=0$) for ladders with $\Delta>0$ described by the Hamiltonian in Eq.~\eqref{eq:XX_ladder_ham}. The parameters for this study were determined as discussed in Sec.~\ref{sec:model}.

\begin{figure}
\includegraphics[width=\linewidth]{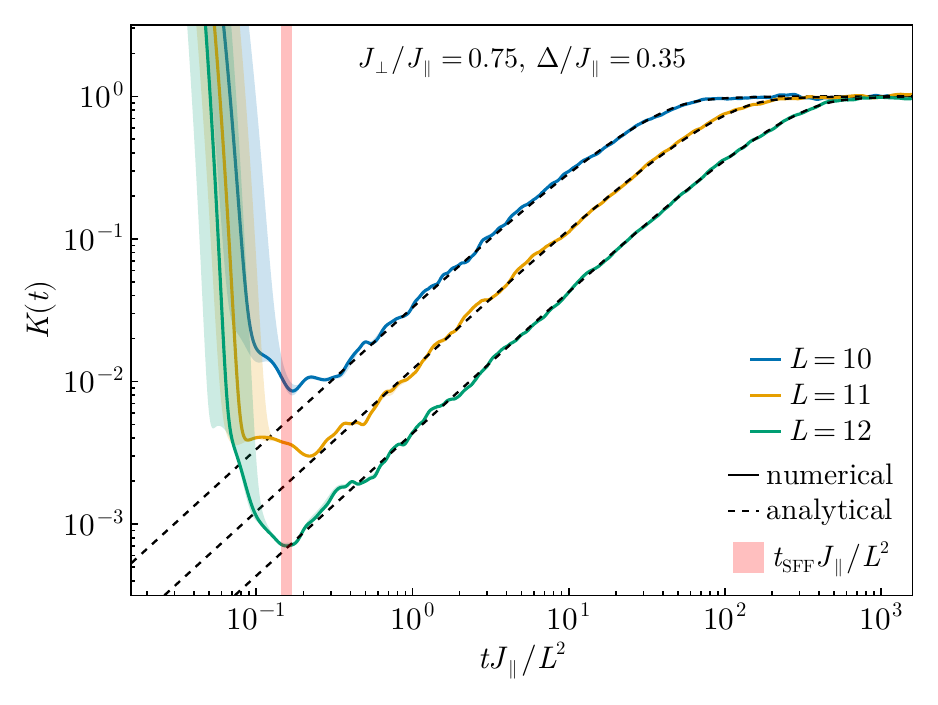}
\vspace{-0.7cm}
\caption{SFF $K(t)$ of clean spin-$\tfrac{1}{2}$ XXZ ladders versus the normalized time for $J_\perp/J_\parallel = 0.75$ and $\Delta / J_\parallel = 0.35$. The solid lines show the numerical results for the SFFs (obtained as weighted averages over symmetry sectors) for $\eta = 0.3$ and the dashed lines show the analytical GOE prediction $K_\mathrm{GOE}(t)$. The shaded bands around the SFF results, mainly visible in the nonuniversal short-time regime, show the range of SFF curves obtained for $\eta \in [0.2, 0.4]$. The vertical shaded regions show the range of SFF times $t^{}_\mathrm{SFF}$, extracted from the largest-accessible (linear) ladder size $L=12$ using $\eta\in[0.2, 0.4]$ and $\epsilon^{}_{\Delta K} \in [0.01, 0.06]$, further widened to take into account the uncertainty in the Heisenberg times. See Appendix~\ref{sec:sff} for details.}
    \label{fig:sff_vs_t_clean}
\end{figure}

\begin{figure}
\includegraphics[width=\linewidth]{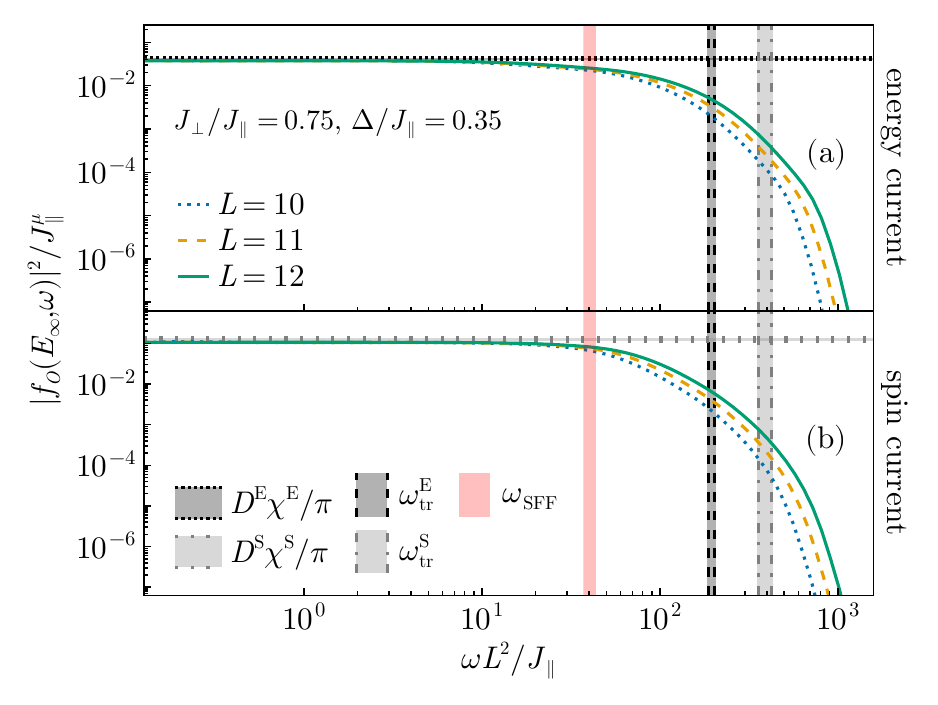}
\vspace{-0.7cm}
\caption{Spectral function for a clean ladder model with an additional Ising interaction on the legs [$J_\perp/J_\parallel = 0.75$, $\Delta / J_\parallel = 0.35$, see Eq.~\eqref{eq:Delta}], plotted as $\abs{f^{}_O(E_\infty , \omega)}^2$ versus $\omega L^2 /  J_\parallel$ for (a) the energy current and (b) the spin current, for $L = 10, 11, 12$ (see the legend). The vertical and horizontal shaded bands are as in Fig.~\ref{fig:off-diag_ETH_vs_omega_L_squared}. See the text and Appendixes for further details and simulation parameters.}
\label{fig:off-diag_ETH_vs_omega_L_squared_clean}
\end{figure}

In Fig.~\ref{fig:sff_vs_t_clean}, we show the SFFs obtained as weighted averages over symmetry sectors. Remarkably, this averaging is sufficient to make the RMT ramp for $t^{}_{\mathrm{SFF}}< t< t^{}_{\mathrm{H}}$ and the long-time saturation at $t = t^{}_{\mathrm{H}}$ clearly visible. There are, however, still visible fluctuations of the SFF on top of the RMT ramps and plateaus. We find that those fluctuations decrease with increasing system size, i.e.,~as the number of symmetry sectors increases. Therefore, these fluctuations may disappear for sufficiently large system sizes, in which case averaging over symmetry sectors may play an equivalent role to averaging over disorder realizations in the disordered case. We extract $t^{}_\mathrm{SFF}$ from the largest available (linear) system size, $L=12$, for $\eta\in [0.2, 0.4]$ and $\epsilon^{}_{\Delta K}\in [0.01, 0.06]$. The shaded band shows the range of values obtained, further widened to take into account the uncertainty in the Heisenberg time (see Appendix~\ref{sec:sff}).

In Fig.~\ref{fig:off-diag_ETH_vs_omega_L_squared_clean}, we show our results for the spectral function in the clean spin-$\tfrac{1}{2}$ XXZ ladders. They exhibit clear plateaus at system-size-independent values that are consistent with the predictions of Eq.~\eqref{eq:fzero}. Like in the disordered ladders, we find that the vertical shaded regions, i.e., the estimated range of values of $\omega_{\mathrm{SFF}}^{}$, appear to mark the end of the plateau, i.e.,~within our uncertainties in the determination of $\omega^{}_{\mathrm{SFF}}$, we find it to be consistent with $\omega^{}_{\mathrm{ETH}}$. We also find $\omega_{\mathrm{tr}}^{\mathrm{E}}$ and $\omega_{\mathrm{tr}}^{\mathrm{S}}$ to be larger than $\omega^{}_{\mathrm{SFF}}$, so the relations in Eq.~\eqref{eq:relationsbetweenomegas} also apply to the clean case.

\section{Summary and outlook}\label{sec:conclusions}

We studied the spectral functions of the total energy and spin currents for disordered and clean spin ladders in the ETH regime. We found that, at low frequencies $\omega \lesssim \omega^{}_{\mathrm{ETH}}$, these spectral functions exhibit plateaus at system-size-independent values that are consistent with the predictions of Eq.~\eqref{eq:fzero}. The plateaus in the spectral functions are expected to occur below a frequency $\omega^{}_{\mathrm{ETH}}$, a regime in which the matrix elements of observables are expected to exhibit RMT statistics. In general, there is no systematic way to extract $\omega^{}_{\mathrm{ETH}}$ from the spectral function, but for the spin-$\tfrac12$ ladders considered here, we find that $\omega^{}_{\mathrm{ETH}}$ is consistent with $\omega^{}_{\mathrm{SFF}}$ extracted from the SFF. The transport frequencies, using the definition $\omega^{\mathrm{ch}}_{\mathrm{tr}} = 2\pi (2\pi /L)^2 D^{\mathrm{ch}}$, on the other hand, were found to be larger than $\omega^{}_{\mathrm{ETH}}$. Thus, there is a need to gain a better understanding of the non-universal prefactors in these quantities. A potentially related question that is left for future studies is how the extent of the plateaus in the spectral functions depends on the observable considered. In this regard it will be important to consider observables that overlap with different powers of the Hamiltonian and/or other conserved quantities such as the magnetization.

Another important result of our study is the finding that averaging the SFFs calculated in different symmetry sectors of a translationally invariant Hamiltonian suppresses fluctuations and allows one to resolve the RMT ramp in the SFF. In translationally invariant Hamiltonians, the number of symmetry sectors (labeled, among other quantum numbers, by the total quasimomentum) is proportional to the number of lattice sites so it increases with increasing system size. It will be interesting to explore in the future the role that the number of lattice sites has in the fluctuations remaining after the average. As another outlook, we remark that the current main limitation in accessing larger systems derives from the need to use exact diagonalization or other finite-system size methods~\cite{Sierant2020A} to compute the spectral function and the spectral form factor. A potentially promising route would be to use recently developed matrix-product-states methods with filters to study larger system sizes~\cite{Luo2024}.

\begin{acknowledgments}
We thank Marcin Mierzejewski, Anatoli Polkovnikov, and Rafał \'{S}wi\c{e}tek for helpful discussions. This work was funded by the Deutsche Forschungsgemeinschaft (DFG, German Research Foundation)---499180199, 493420525, 436382789, 405797229, via FOR 5522 and large equipment grants (GOEGrid cluster, Scientific Compute Cluster at GWDG), and by the United States National Science Foundation (NSF) Grant No.~PHY-2309146 (R.P.~and M.R.). Parts of the computations were performed on the Scientific Compute Cluster at GWDG---the joint data center of the Max Planck Society (MPG) and Georg-August-Universität Göttingen.
\end{acknowledgments}

\section*{Data \texorpdfstring{\&}{&} code availability}
The data that support the findings of this article are partially
openly available~\cite{zenodo}.

\appendix

\section{Calculation of the SFF}\label{sec:sff}

We compute the SFF using the following equation:
\begin{equation}\label{eq:sff}
    K{(\tau)} =  \frac{\abs{\sum_{\alpha} \rho^{}_\mathrm{f}{\qty(\varepsilon^{}_\alpha; \eta)} \exp(-i \varepsilon^{}_\alpha \tau)}^2}{\sum_\alpha \abs{\rho^{}_\mathrm{f}(\varepsilon^{}_\alpha; \eta)}^2}\,,
\end{equation}
where $\varepsilon^{}_\alpha$ denotes the unfolded energy eigenvalues for a given disorder realization or in a given symmetry sector and $\tau = 2\pi t / t^{}_\mathrm{H}$ is the normalized time, while $t^{}_\mathrm{H}$ is the Heisenberg time. The unfolding procedure fixes the mean-level spacing to unity and thus the normalized Heisenberg time to $\tau^{}_\mathrm{H}=2\pi$. $\rho^{}_\mathrm{f}(\varepsilon^{}_\alpha; \eta)=\exp[-\tfrac{(\varepsilon^{}_\alpha-\bar{\varepsilon})^2}{2\eta^2\upsigma_\varepsilon^2}]$, where $\bar{\varepsilon}$ and $\upsigma^{}_\varepsilon$ are the mean and standard deviation of the unfolded spectrum, respectively, and $\eta$ controls the width of the Gaussian filter. For the disordered case, we carry out an average over disorder realizations, and for the clean case we average over symmetry sectors. We additionally perform a running average in time to reduce temporal fluctuations. The SFF is computed with a discrete step $\Delta(\log_{10}\tau)=0.001$ and, both to plot the SFF in the figures and to determine $t^{}_{\mathrm{SFF}}$, we use a running average window on a $\log_{10}$ scale, $\delta(\log_{10}\tau)=0.025$. To lighten the notation, we do not indicate these averages with additional symbols.

To extract $t^{}_\mathrm{SFF}$, we first work with normalized times $\tau^{}_{\mathrm{SFF}}=2\pi t^{}_{\mathrm{SFF}}/t^{}_\mathrm{H}$ to compare the numerically obtained SFF to the RMT prediction for the Gaussian orthogonal ensemble (GOE). We evaluate the criterion function~\cite{Suntajs2020}
\begin{equation}
    \Delta K(\tau) = \abs{\log_{10}\frac{K(\tau)}{K_\mathrm{GOE}(\tau)}}\,,
\end{equation}
where $K_\mathrm{GOE}(\tau)$ is the analytical GOE result for the SFF. $K_\mathrm{GOE}(\tau)$ is given by:
\begin{equation}
    K_\mathrm{GOE}{(\tau)} = \begin{cases}
         \frac{\tau}{\pi} - \frac{\tau}{2\pi}\ln(\frac{\tau}{\pi} + 1) & \qif* \tau \leq 2\pi\\
         2 -\frac{\tau}{2\pi} \ln(\frac{\frac{\tau}{\pi} + 1}{\frac{\tau}{\pi} - 1}) & \qif* \tau > 2\pi
    \end{cases}\,.
\end{equation}

We then determine $\tau^{}_\mathrm{SFF}$ as the (normalized) time at which
\begin{equation}
    \Delta K(\tau^{}_\mathrm{SFF}) = \epsilon^{}_{\Delta K}\,, \label{eq:crit}
\end{equation}
with $\epsilon^{}_{\Delta K}$ a chosen threshold. Since we work with discrete data points, we use the time point at which $ \Delta K(\tau^{}_\mathrm{SFF})$ is below the threshold for the first time as our estimate of $\tau^{}_\mathrm{SFF}$. The chosen discretization yields an uncertainty of less than 2\% in the results for $t_{\mathrm{SFF}}$ for a given pair of $(\eta, \epsilon_{\Delta K})$.

To estimate the systematic uncertainty in the extraction of $t^{}_\mathrm{SFF}$, we find a range of values of $\eta$ for which the results for $t^{}_\mathrm{SFF}$ are insensitive to the specific value of $\eta$ chosen. For any given value of $\eta$, we vary $\epsilon^{}_{\Delta K}$ to show how sensitive $t^{}_\mathrm{SFF}$ is to the criterion chosen. We always identify $\tau^{}_\mathrm{SFF}$ as the earliest time at which the deviation $\Delta K(\tau)$ comes within $\epsilon^{}_{\Delta K}$ according to Eq.~\eqref{eq:crit}. Scanning from early times makes the estimate insensitive to the fluctuations of $\Delta K(\tau)$ beyond $\tau^{}_\mathrm{SFF}$. It also means that the larger the fluctuations of $\Delta K(\tau)$, the larger the values one should select for $\epsilon^{}_{\Delta K}$.

For our disordered and clean ladders, we find that for $\eta \in [0.2,0.4]$, $t_{\mathrm{SFF}}$ exhibits only a very weak dependence on $\eta$ across all models and system sizes studied. For the disordered ladders, for which the fluctuations about the RMT ramp are very small, we select $\epsilon^{}_{\Delta K} \in [0.005,0.025]$. For the clean ladders, for which larger fluctuations are observed, we select $\epsilon^{}_{\Delta K} \in [0.01,0.06]$. The results reported for $t_{\mathrm{SFF}}$ in the main text are presented as an interval obtained for these input parameter ranges for $(\eta,\epsilon^{}_{\Delta K})$.

\section{Calculation of the spectral function}\label{sec:cal_f_O}

The spectral functions are calculated using full exact diagonalization as follows.

\paragraph{Matrix elements within a symmetry sector} For a given set of model parameters (and, in the disordered case, for a given disorder realization), we diagonalize $\qop{H}$ separately within each resolved symmetry sector labeled by a set of quantum numbers $\{ q \}$, to obtain the energy eigenstates $\{\ket{\psi_\alpha}\}$ in that sector. In the disordered case the only conserved quantity is the total magnetization and we only consider the zero-magnetization sector, $S^z =0$; in the clean case, we additionally have total quasimomentum $k^{}_x$, the quantum number $p^{\mathrm{r}}_y$ associated with reflection symmetry along the rungs, and $p^{\mathrm{S}}_z$ associated with the spin-$z$ parity symmetry. We then evaluate 
\begin{equation}
    O_{\alpha \beta} = \mel{\psi_\alpha}{\qop{O}}{\psi_\beta}\,, \qquad \alpha \neq \beta
\end{equation}
for the current operators $\qop{O} = \qop{J}^{\mathrm{ch}}$, with $\mathrm{ch} = \mathrm{S}, \mathrm{E}$. Associated with the matrix elements $O_{\alpha \beta}$, we label the mean energies as $\bar{E}_{\alpha\beta} = [E_\alpha + E_\beta]/2$ and frequencies as $\omega^{}_{\alpha\beta} = E_\alpha - E_\beta$.

\paragraph{Infinite-temperature average} We evaluate the spectral function at infinite temperature, i.e.,~at the (sector-resolved) mean energy
\begin{equation}
    E_\infty(\{ q \}) \equiv \langle\qop{H}\rangle_{\{ q \}} = \frac{\Tr_{\{ q \}}{\qop{H}}}{\mathcal{D}(\{ q \})}\,.
\end{equation}
Namely, in our averages of $|O_{\alpha \beta}|^2$ used to compute the spectral function, we only include pairs of energy eigenstates satisfying:
\begin{equation}\label{eq:supp_energy_window}
    \abs{\bar{E}_{\alpha\beta} - E_\infty(\{ q \})} \leq \Delta E \sqrt{2L}\,,
\end{equation}
i.e.,~within an energy window of half width $\Delta E \sqrt{2L}$ around the infinite-temperature mean energy of sector $\{ q \}$. We use $\Delta E = 0.025 J_\parallel$ throughout this work.

\paragraph{Density of states} The density of states $\Omega(E_\infty, \Delta E)$ is evaluated per sector as
\begin{equation}
    \Omega(E_\infty, \Delta E; \{ q \}) \!=\! \frac{\#\{ \alpha \!:\! \abs{E_\alpha(\{ q \}) \! -\! E_\infty(\{ q \})}\! \leq\! \Delta E \sqrt{2L} \}}{2 \Delta E \sqrt{2L}},
\end{equation}
i.e.,~the number of eigenstates whose energy lies in the window in Eq.~\eqref{eq:supp_energy_window}, divided by the window width.

\paragraph{Frequency binning and moving average} To compute the frequency dependence of the infinite-temperature spectral function, the calculated values of $\abs{O_{\alpha\beta}}^2$ are first averaged within logarithmically spaced frequency bins and then smoothed via a centered moving average,
\begin{equation}
    \eval{\overline{\abs{O_{\alpha\beta}}^2}}_{\omega; \{ q \}} = \frac{1}{2M_\omega+1}\!\!\sum_{n=-M_\omega}^{M_\omega} \!\!\!\abs{O_{\alpha\beta}}^2\Big|_{\log_{10}\!\!\qty[\frac{|\omega^{}_{\alpha\beta}|}{J_\parallel}]\! \in \mathrm{bin}(n)},
\end{equation}
using a logarithmic bin width $\Delta(\log_{10}[|\omega|/J_\parallel]) = 0.05$ and $2M_\omega + 1 = 7$ bins per window ($M_\omega = 3$, corresponding to three bins on either side of the central one).

\paragraph{Spectral function and averages} Combining the numerical results for density of states and for the smoothed matrix elements gives the symmetry-sector resolved spectral function, 
\begin{equation}
    \abs{f^{}_O(E_\infty, \omega;\{q\}) }^2 \approx \Omega(E_\infty, \Delta E; \{ q \}) \, \eval{\overline{\abs{O_{\alpha \beta}}^2}}_{\omega; \{ q \}}.
\end{equation}

\begin{figure}
\includegraphics[width=\linewidth]{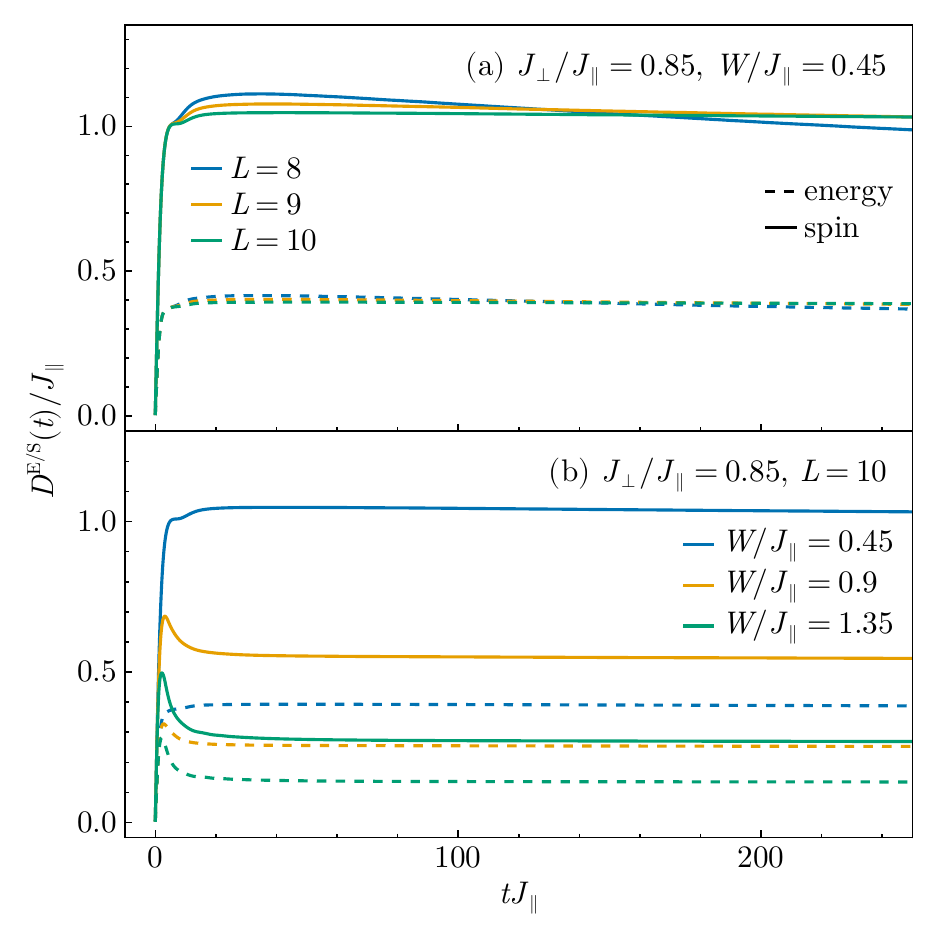}
\vspace{-0.8cm}
\caption{Time-dependent diffusion constants $D^\mathrm{ch}(t)$, $\mathrm{ch}=\mathrm{E,S}$ of a spin-$\tfrac12$ XX ladder as a function of time in units of  $1/J_\parallel$, for $J_\perp/J_\parallel = 0.85$. (a) $W/J_\parallel = 0.45$ for different ladder sizes. (b) $L=10$, for different disorder strengths. In both subplots, the dashed lines represent the energy-diffusion constant ${D^\mathrm{E}}(t)$, while the solid lines denote the spin-diffusion constant ${D^\mathrm{S}}(t)$. Each diffusion constant is averaged over $M=500$ disorder realizations. The time evolution is performed using the Lanczos algorithm with a discrete time step of $\Delta t =0.01/J_\parallel$ and a maximum of $20$ Lanczos iterations. }
\label{fig:D_vs_t}
\end{figure}

In the clean case, we average over the symmetry sectors $\{ q \}$ within the zero-magnetization sector, weighting each sector by its Hilbert-space dimension $\mathcal{D}(\{ q \})$,
\begin{equation}
    \abs{f^{}_O(E_\infty, \omega)}^2 \approx \frac{\sum^\prime_{\{ q \}} \mathcal{D}(\{ q \}) \, \abs{f^{}_O(E_\infty, \omega; \{ q \})}^2}{\sum^\prime_{\{ q \}} \mathcal{D}(\{ q \})},
\end{equation}
where the primed sum excludes the total quasimomentum sectors $k^{}_x =0,\pi$, which exhibit an additional parity symmetry along the legs and stronger finite-size effects~\cite{LeBlond2020}.

In the disordered case, the spectral function is computed independently for each disorder realization, and the final result shown in Fig.~\ref{fig:off-diag_ETH_vs_omega_L_squared} is the average over $M$ realizations.

\section{Diffusion constant calculations}\label{sec:diff_const}

\begin{figure}
\includegraphics[width=\linewidth]{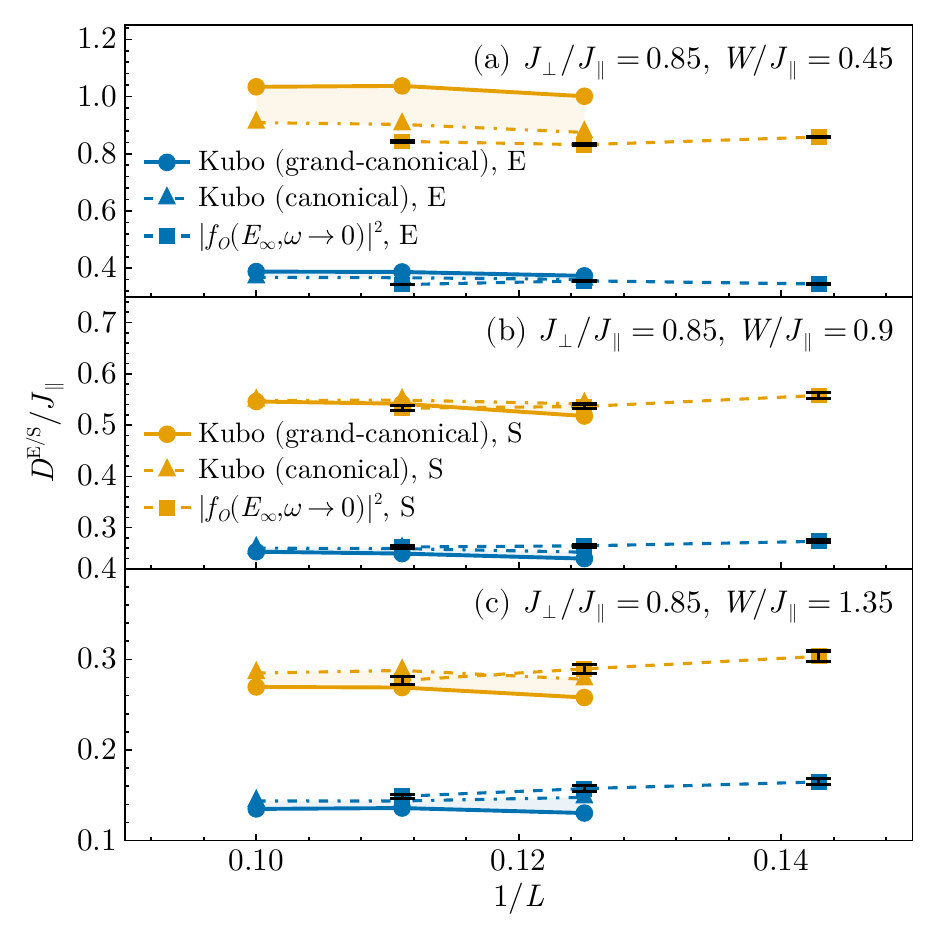}
\vspace{-0.8cm}
\caption{Averaged diffusion constants $D^\mathrm{E}(L)$ and $D^\mathrm{S}(L)$ of disordered spin-$\tfrac12$ XX ladders versus $1/L$ for $J_\perp/J_\parallel = 0.85$ and disorder strength (a) $W/J_\parallel = 0.45$, (b) $W/J_\parallel = 0.9$, and (c) $W/J_\parallel = 1.35$. Blue (orange) curves show the energy (spin) diffusion constants. Solid lines with circles show the grand-canonical Kubo estimates $D^\mathrm{E}(L)$ and $D^\mathrm{S}(L)$ from the current autocorrelation function, averaged over $200 < t J_\parallel < 250$ for $L = 8,9,10$, with error bars indicating the standard deviation of $D^{\mathrm{E}}(t)$ and $D^\mathrm{S}(t)$ within this window. Dash-dotted lines with triangles show the corresponding canonical (i.e.,~fixed $S^z=0$) Kubo estimates; the shaded region between the grand-canonical and canonical curves indicates the ensemble dependence. Dashed lines with squares show the diffusion constant estimate from the zero-frequency plateau of the off-diagonal spectral function, $D^\mathrm{ch} = \pi \abs{f^{}_O(E_\infty, \omega \rightarrow 0)}^2 / \chi^\mathrm{ch}$ for $L=7,8,9$; the error bars show the standard deviation of the values across disorder realizations.}
\label{fig:D_vs_inverse_L_disorder}
\end{figure}

We first discuss the disordered case. We show typical results for the time-dependent disorder-averaged diffusion constants for both energy and spin, $D^{\mathrm{E}}(t)$ and $D^{\mathrm{S}}(t)$, in Fig.~\ref{fig:D_vs_t}(a) as dashed and solid lines, respectively. The data are plotted as functions of time in units of $1/J_\parallel$ for the parameters of Fig.~\ref{fig:off-diag_ETH_vs_omega_L_squared}(a) for $L = 8, 9, 10$.

\begin{figure}
\includegraphics[width=\linewidth]{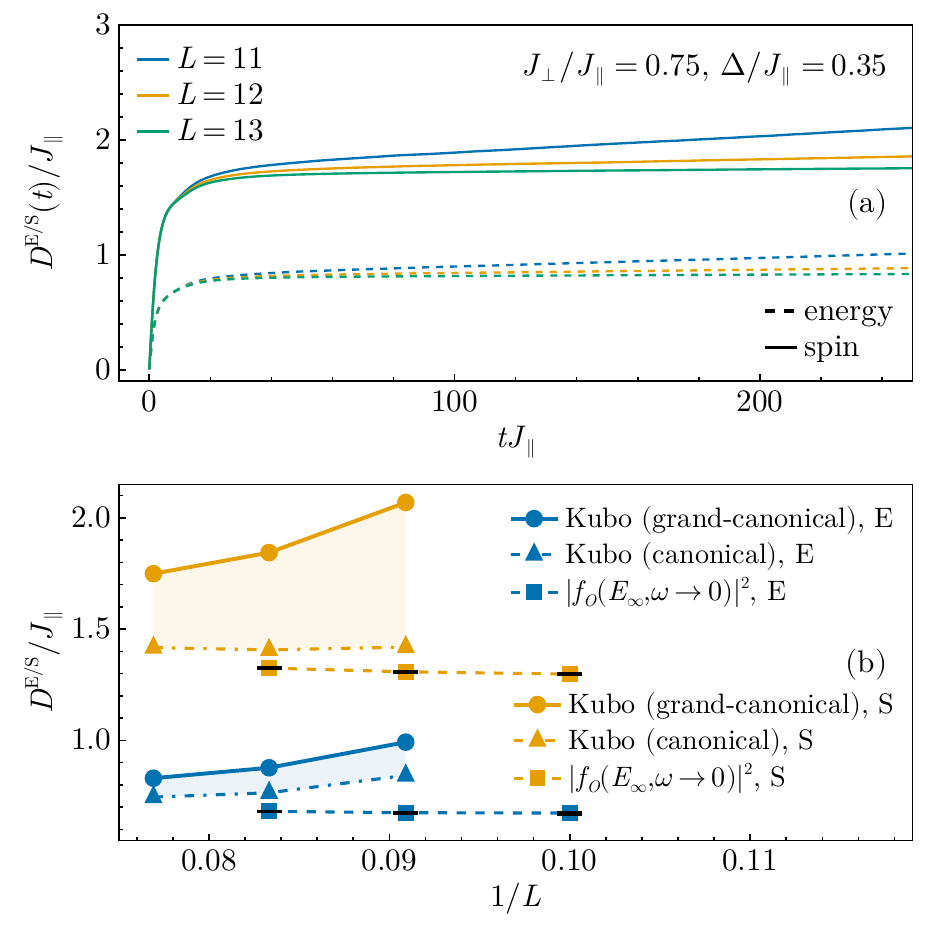}
\vspace{-0.8cm}
\caption{Diffusion constants of clean spin-$\tfrac{1}{2}$ XXZ ladders for $J_\perp/J_\parallel = 0.75$ and $\Delta/J_\parallel=0.35$. (a) Time-dependent diffusion constants $D^\mathrm{E}(t)$ and $D^\mathrm{S}(t)$ vs $t J_\parallel$ for $L \in [11,13]$, from Lanczos time evolution with $\Delta t = 0.01/J_\parallel$ and $20$ Lanczos iterations. The dashed and solid lines show $D^\mathrm{E}(t)$ and $D^\mathrm{S}(t)$, respectively. (b) Averaged diffusion constants $D^\mathrm{E}(L)$ and $D^\mathrm{S}(L)$ vs $1/L$, with blue and orange colors denoting the energy and spin channels. Solid lines with circles show the grand-canonical Kubo estimates, averaged over $200 < t J_\parallel < 250$, with error bars from the standard deviation within this window. Dash-dotted lines with triangles show the canonical (i.e.,~$S^z = 0$) Kubo estimates, and the shaded region indicates the ensemble dependence between these two. Dashed lines with squares show the plateau estimate from the off-diagonal function, $D^{\mathrm{ch}} = \pi \abs{f^{}_O(E_\infty, \omega \rightarrow 0)}^2 / \chi^\mathrm{ch}$. The error bars show the standard deviation of the values across symmetry sectors.}
\label{fig:D_vs_t_and_D_vs_inverse_L_clean}
\end{figure}

As $L$ and time increase, both diffusion constants $D^\mathrm{E}(t)$ and $D^\mathrm{S}(t)$ settle to constant values, with small finite-size dependencies at long times remaining. Note that for this model and choice of parameters, $D^\mathrm{S}>D^\mathrm{E}$ as discussed in Ref.~\cite{Ceven2026}. Figure~\ref{fig:D_vs_t}(b) shows data for fixed $L$ but different disorder strength. In all cases, the diffusion constants decrease with increasing $W$, as expected, yet remain nonzero.

To estimate the large-system diffusion constant and its uncertainty, we proceed as follows. For a fixed $L$, we use the time interval $t J_\parallel \in [200, 250]$ and average $D^\mathrm{ch}(t)$ over time to obtain $D^{\mathrm{ch}}(L)$. This procedure is carried out both in the grand-canonical and the canonical (only considering the $S^z=0$ sector) ensembles. The resulting data are shown in Fig.~\ref{fig:D_vs_inverse_L_disorder}. These different data sets typically differ, allowing us to estimate a range for the thermodynamic-limit diffusion constants. We use the range spanned by the canonical and grand-canonical results for the largest system size in the main figures to indicate the location of the transport times and frequencies. The difference between the two estimates is a conservative estimate of the uncertainty. Note that we also include the diffusion constants from the zero-frequency limit of the corresponding spectral functions in Fig.~\ref{fig:D_vs_inverse_L_disorder}. They are all comparable within the known systematic limitations in calculations of diffusion constants (e.g.,~finite system size and  accessible time scales)~\cite{Bertini2021}.

For the clean ladders, see Fig.~\ref{fig:D_vs_t_and_D_vs_inverse_L_clean}(a), the time-dependent diffusion constant grows rapidly at short times and then crosses over to a slow long-time increase due to the finite-size Drude weight~\cite{Steinigeweg2014A, Steinigeweg2014C}. Since the Drude weight of a nonintegrable model that obeys ETH decays exponentially with $L$~\cite{Steinigeweg2013}, we can use an average at long times and extract $D^\mathrm{E}$ and $D^\mathrm{S}$ in the same manner as in the disordered case. The comparison of data computed for the canonical and grand-canonical ensembles is shown in Fig.~\ref{fig:D_vs_t_and_D_vs_inverse_L_clean}(b). The finite-size dependence is similar to the disordered case and, hence, we estimate diffusion constants and errors analogously.

\bibliography{references}

\end{document}